# Why and How Did the COVID Pandemic End Abruptly?

J. C. Phillips

Dept. of Physics and Astronomy

Rutgers University, Piscataway, N. J., 08854

Abstract

Phase transition theory, implemented quantitatively by thermodynamic scaling, has explained the evolution of Coronavirus' extremely high contagiousness caused by a few key mutations from CoV2003 to CoV2019, identified among 1200 Spike amino acids, as well as the later 2021 evolution to Omicron caused by 30 mutations. It also showed that the Post-Omicron 2022 strain BA.5 with few mutations began a new path. The new path emphasizes enhanced attachment rates in place of former domain synchronization. Here we show that the Post-Omicron early 2023 strains BKK with one stiffening mutation confirm that path, and the single flexing mutation of a later 2023 variant EG.5 strengthens it further. The Post-Omicron path has greatly reduced pandemic deaths, for mechanical reasons involved in amino acid - water interactions, described here very accurately.

Introduction

According to the Worldometer [1], in the Fall and Winter of 2020-2021 (from Nov. 17, 2020, until Feb. 20, 2021) the total deaths in the United States from COVID increased by 250,00. In the Fall and Winter of 2021-2022 (from Aug. 21, 2021, until Mar. 3, 2022) there was another increase of 1/3 to 330,000. Then from Aug. 21, 2022 until Mar. 12, 2023, the number decreased abruptly by 3/4, falling 330,000 to only 80,000. The abrupt decrease was not caused by the amazing vaccine that had been widely available after Spring 2021. The rapid decrease after summer 2022 suggests a possible phase transition in the molecular structure of COVID spikes. Here we show a phase-transition based explanation for this abrupt decrease, based on earlier papers that analyzed the 1200 - COVID Spike amino acid sequences from 2003 until BA.5, 2022 [2-7].

Methods

The methods used in this paper are drawn from a wide range of disciplines. These have been discussed in our earlier articles [2-7]. While phase transitions are a familiar part of daily life (boiling or freezing water), the methods of statistical mechanics involving phase transitions are well known mainly to chemists and physicists [8]. Note that the CoV new path was identified in [7] using no new parameters in phase transition theory. Our methods are based on conjectures in physics concerning the evolution of self-organized networks [8], which can include even inorganic network glasses, which have been selected to optimize their functional properties. This conjecture has attracted much interest, and it is summarized in the abstract of [8], which should be read as part of this paper. (Note: [8] was in the upper 1% in citations of physics papers in 2018.) Here the optimization of CoV evolution has involved increasing contagiousness.

Results

The spike (S) protein of SARS-CoV-2, which plays a key role in the receptor recognition and cell membrane fusion process, is long and slender, and immersed in water. It is composed of two subunits, S1 and S2. The S1 subunit contains a Receptor Binding Domain (RBD) that recognizes and binds to the host receptor angiotensin-converting enzyme 2, while the S2 subunit mediates viral cell membrane fusion. Spike dynamics were explained [4-7] in terms of domain motions synchronized through leveling of hydrophilic and/ or hydrophobic extrema in sliding hydropathic windows. The key extrema are in or near the RBD. These extrema were shown to correlate well with COVID evolution only when interactions of amino acids with water were averaged over sliding windows of lengths W ~ 35-39 amino acids. Thus the common practice of focusing on individual mutations (W = 1 or 3, including nearest neighbors) has not succeeded in explaining COVID evolution. The RBD is L ~ 70 amino acids wide, so it is not surprising that the effects of domain synchronization are best resolved with W ~ 35-39 ~ L/2.

There was a large increase in extrema leveling from 2003 to 2019, resulting in greater contagiousness, in accordance with natural selection. This long-term trend was abruptly reversed with the new 2022 variant BA.5, described predictively as a "New Twist" in Aug. 2022 [7]. Specifically the largest changes in 2022 were caused by a few mutations that increased two hydrophobic pivots (numbered 7 and 8, centered at $369 \pm 6$ for $W \pm 10$, and $491 \pm 6$ for $W \pm 10$) and one (number 1, centered at $448 \pm 5$ for $W \pm 10$) hydrophilic flexing (W = 39, as in [5]) extrema in the ~ 70 amino acid RCB (see Fig. 1, copied here for the reader's convenience). These changes were largest in edge 8, which increased from 169.4 (Omicron) to 171.9 (BA.2) to 172.4 (BA.5). This trend was continued in early 2023 strains labeled XBB, with one mutation S486P [9]. This single mutation increased edge 8 to 172.9, in excellent agreement with the trend of edge 8 identified earlier. Finally the single mutation in late 2023, F456L of EG.5, deepens the hydrophilic edge 1, from 138.5 in Omicron, to 138.0 (see Table 1 of [7]). Note that the two new mutations are less than 7 sites away from their respective centers, and thus not only lie well within the $W = 39$ sliding window; they would have been near the centers for any W as small as 15.

While this double success is gratifying, how likely is it to be only accidental? A $W = 15$ window associated with extremum 8 is the fraction ∼ 15/1200 = 5/400 of the Spike sequence. Also the mutation should increase an average value of 172, and this occurs on average 0.4. Thus the odds against this success happening accidentally are more than 100 to 1. The second success is equally unlikely, so the odds against the two combined successes are more than 10,000 to 1. Moreover, Prolines stiffen proteins mechanically, because Pro is the only amino acid which is connected twice to the peptide backbone [7]. The fraction of Prolines in all proteins is around 6%, and in Spikes only 4.4%. However, in window 8 of BA.5 it is already 10%. Adding one more Pro in XBB brings this fraction to 12%, nearly three times that of the entire Spike. Natural selection has redirected Spike evolution with < 0.6 % mutations, resulting in the abrupt drop in deaths.

It was suggested in [7] that increases in a few extrema in the RBD could increase contagiousness by increasing viral attachment in the upper respiratory tract in S1. However, cell fusion through S2 involves an intermediate state in which the binding of S1 to the substrate is weakened, see Fig.1 of [10]. This is less likely to occur in BA.5 than in BA.2 [7]. The stiffening effect of the mutation S486P in XBB strains further strengthens binding to the RBD, but what of the flexing effect in EG.5 with F456L? Here we see in Fig. 1 that the deeply hydrophilic edge 1 separates the two hydrophobic edges 7 and 8. Thus edge 1 can act as a hinge speeding cooperative pinning by edges 7 and 8 during the transition state of the Spike to tissue. Thus combined hydrophobic and hydrophilic mechanical stabilization of binding in the upper respiratory tract is a natural explanation for the abrupt decrease in deaths, as BA.5 spread rapidly in the United States in Fall 2022 [11].

Discussion

The successes of hydropathic methods (against odds of >10,000 to 1) in describing the BKK and EG.5 single mutations may be surprising. This is less so when one realizes that these successes emerged from general theories regarding the tendency of self-organized networks to evolve towards criticality and thus approach phase transitions [8]. The input to the theory of large-scale second-order phase transitions is based on a hydropathicity scale developed from studying the W-dependent amino acid contributions to the curvatures of > 5000 protein segments [12]. The successes described here in connecting this structural data to pandemic evolution have occurred through combined international [8,12] and interdisciplinary efforts, some of which are listed in the 57 references of [4]. The present calculations are not elaborate; they involved only a desk-top computer and an EXCEL module. The simplicity is made possible by the critical action of water in shaping dynamical domains.

What does the future hold? It seems unlikely that natural selection will repeat domain synchronization by leveling of extrema, as it did in 2003 - 2022. The XBB and EG.5 single mutations strengthened edges 8 and 1, perhaps the next single mutation will strengthen edge 7. We expect tightening of binding to the RBD and further decreases in the US death rate, perhaps to 60,000/year, compared to 50,000/year from flu. Many mechanisms have been proposed that might explain why the 1918 flu was so deadly [13], and that strain has disappeared.

An alternative study focused on a few mutations near the S1/S2 cleavage site ~ 685 [14], which might have explained the abrupt decrease in the death rate from Omicron to BA.5. However, the quantitative features identified here are associated with improvements in binding in the RBD (437-508), far from the S1/S2 cleavage. Theoretical research on effects of protein single mutations has concentrated on protein structural stability [15], as changes in functionality are not easily treated by other methods. The predictive features of the present method suggest that science has advanced greatly in its ability to understand protein evolution. There are interesting similarities between the evolving elastic properties of Spikes and very nearly homogeneous covalent glass alloys [16].

## References

1. Worldometer, https://www.worldometers.info/coronavirus/country/us/ (2023).
2. Phillips, J. C. Scaling and self-organized criticality in proteins: Lysozyme *c*. Phys. Rev. E **80**, 051916 (2009)

3. Phillips, J.C. Fractals and self-organized criticality in proteins. Phys A **415**, 440-448 (2014).

4. Phillips, JC Synchronized attachment and the Darwinian evolution of Coronaviruses CoV-1 and CoV-2. arXiv2008.12168 (2020); Phys. A **581**, 126202 (2021).

5. Phillips, JC Moret, MA Zebende, G Chow, CC Phase transitions may explain why SARS-CoV-2 spreads so fast and why new variants are spreading faster. Phys. A **598**, 127318 (2022).

6. Phillips, JC What Omicron does, and how it does it. arXiv 2112.04915 (2021).

7. Phillips, JC From Omicron to Its BA.5 Subvariant: A New Twist in Coronavirus Contagiousness Evolution. arXiv 2208.04956 (2022).

8. Munoz, M. A. Colloquium: Criticality and dynamical scaling in living systems. Rev. Mod. Phys. **90**, 031001 (2018).

9. Zappa M, Verdecchia P, Angeli F. Severe acute respiratory syndrome coronavirus 2 evolution: How mutations affect XBB.1.5 variant. Eur J Intern Med. 2023 Jun;112:128-132.

10. Huang, Y., Yang, C., Xu, Xf. *et al.* Structural and functional properties of SARS-CoV-2 spike protein: potential antivirus drug development for COVID-19. *Acta Pharmacol Sin* **41**, 1141–1149 (2020).

11. Am Med. Assoc., https://www.ama-assn.org/delivering-care/public-health/omicron-and-ba5-questions-patients-may-have-and-how-answer (Aug., 2022).

12. Moret, M. A.; Zebende, G. F. Amino acid hydrophobicity and accessible surface area. Phys. Rev. E **75**, 011920 (2007).

13. Anon, Why the flu of 1918 was so deadly. https://www.bbc.com/future/article/20181029-why-the-flu-of-1918-was-so-deadly (2018).

14. Park SB, Khan M, Chiliveri SC, Hu X, Irvin P, Leek M, Grieshaber A, Hu Z, Jang ES, Bax A, Liang TJ. SARS-CoV-2 omicron variants harbor spike protein mutations responsible for their attenuated fusogenic phenotype. Commun Biol. 2023 **24**:556.

15. Scarabelli, G Oloo,EO Maier, JKX *et al.* Accurate prediction of protein thermodynamic stability changes upon residue mutation using free energy perturbation. J. Mol. Biol. **434**, 16735 (2022).

16. Chbeir, R Welton, A … Micoulaut, M Boolchand, P Glass transition, topology, and elastic models of Se-based glasses. J. Am. Ceramic Soc. 106, 3277-3302 (2023).

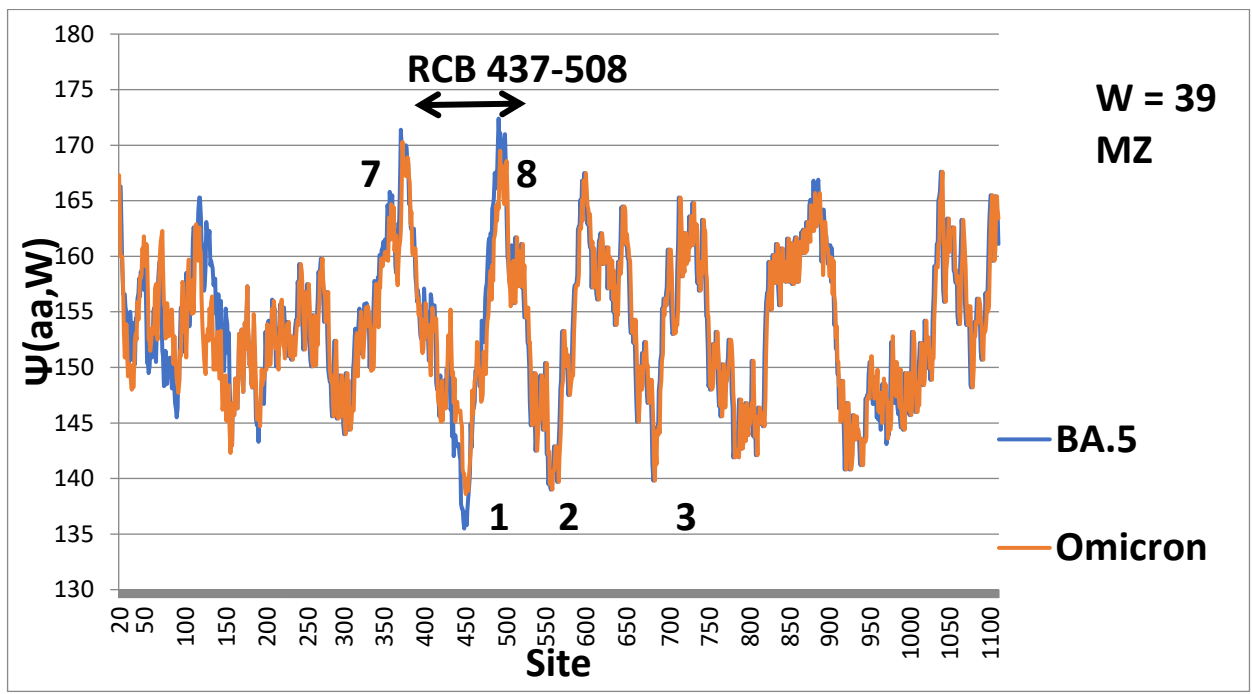

Fig. 1. The hydropathic wave profiles of Omicron and BA.5 are very similar, and the most important changes are in edges 7, 8 and 1. These small changes are systematic, and increase contagiousness. Most recently (since BA.5) they also decrease deaths.